\documentclass[amsmath,amssymb,epsfig,pre,aps]{revtex4}

\usepackage{graphicx}% Include figure files
\usepackage{dcolumn}% Align table columns on decimal point
\usepackage{bm}% bold math

\begin{document}

% Version 1.1  Submitted to PRE 11 July  2003
\date{\today}

\author{I.V. Barashenkov$^\star$}
% \email{igor@cenerentola.mth.uct.ac.za}
\author{S. Cross$^\dag$}
% \email{elena@ultra.jinr.ru}
\affiliation{Department of Mathematics and Applied Mathematics,
University of Cape
Town, Rondebosch 7701, South Africa}
\author{Boris A. Malomed$^\S$}
% malomed@eng.tau.ac.il
\affiliation{Department of Interdisciplinary Studies,
 Faculty of Engineering,
 Tel Aviv University, Tel Aviv 69978, Israel}
\title{Multistable Pulse-like Solutions in a Parametrically Driven
	 Ginzburg-Landau Equation}

 \begin{abstract}
It is well known that pulse-like solutions of the cubic complex
Ginzburg-Landau equation are unstable but can be
stabilised by the addition of  quintic terms. In this paper
we explore an alternative mechanism where
the role of the stabilising agent is played by the
 parametric driver.
Our analysis is based on the numerical
continuation of solutions in one of the parameters of 
the Ginzburg-Landau equation (the diffusion coefficient $c$), starting from
the
 nonlinear Schr\"odinger
limit (for which $c=0$).
The continuation generates, recursively, a sequence of coexisting
stable solutions with increasing number of humps. The sequence
``converges"
to a long pulse which 
 can
be interpreted as a bound state of two 
fronts with opposite polarities. 
\end{abstract}

\maketitle

\section{Introduction}

A variety of nonequilibrium phenomena 
such as open flows in hydrodynamics, thermal
convection in pure fluids and binary mixtures,
 processes
in lasers and nucleation during the 
 first-order phase transitions, 
can be modelled by 
the complex Ginzburg-Landau equations. (For review 
and reference, see e.g. \cite{GL,Baer_Torcini,Akhmediev,Aronson_Kramer,CrossHoh}.)
Of primary importance are pulse-like solitary wave solutions,
which represent localised structures widely observed in nonequilibrium
systems.
It is well known that in the cubic Ginzburg-Landau equation,
\begin{equation}
i \psi_t+ i \gamma \psi +(1-ic) \psi_{xx} + (2-i g) |\psi|^2 \psi= 0,
\label{GLcubic}
\end{equation}
  solitary waves are  unstable for
 all positive $c$, $\gamma$ and real $g$. In the $g>0$ case, however,
they can be stabilised by the addition of quintic terms:
\begin{equation}
i \psi_t+ i \gamma \psi
+(1-ic) \psi_{xx} + (2-i g) |\psi|^2 \psi =-
(q_r+iq_i) |\psi|^4 \psi,
\label{GLquintic}
\end{equation}
with positive
$q_i$ \cite{Borya,Fauve}. This example suggests that
 solitary pulses can  be stable in a more general
class of Ginzburg-Landau equations
 where the zero  solution undergoes a 
subcritical
(rather than supercritical) bifurcation to a flat nonzero
solution
\cite{Fauve}. In  Eq.(\ref{GLquintic}), 
the terms with $\gamma$ and $c$ account for
linear homogeneous and nonhomogeneous losses, respectively,
while the terms with $g$ and $q_i$ describe the cubic gain
and quintic dissipation.

The present work deals with another equation of
the Ginzburg-Landau type
 exhibiting the subcritical bifurcation,
%\cite{Meron,Fauve}, 
viz. the
{\it parametrically driven\/}
 Ginzburg-Landau:
\begin{equation}
i\psi _{t}+i\gamma \psi
+(1-ic)\psi _{xx}+(2-ig) \left| \psi \right|^{2}\psi
=h\psi ^{\ast }e^{2i \omega t}.
\label{eq:ddcgle_bis}
\end{equation}
Here, as in (\ref{GLcubic})-(\ref{GLquintic}), 
the positive $c$ and $\gamma$ are
the homogeneous loss and diffusion coefficients, respectively.
 The term with an asterisk (indicating
 the complex conjugation) represents
 the parametric driver.
 (The driver's amplitude $h$ can be chosen positive without loss of
 generality.)
 When $c=g=0$, Eq.(\ref{eq:ddcgle_bis}) gives the parametrically
 driven damped nonlinear Schr\"odinger equation. 
 This special case has been
 studied extensively;
 in particular, stable solitary waves \cite{BBK} 
 and their stable complexes \cite{BZ} were found,
 and their bifurcations and supercritical dynamics
 analysed \cite{supercritical}. 
  The objective of the present
 work is to advance beyond the nonlinear Schr\"odinger limit. We will still 
 keep $g=0$ but allow for a nonzero diffusion coefficient, $c$. As
 we will see, even such a minimal generalisation gives
 rise to a new phenomenology of localised solutions
 which includes the multistability of pulses and pulse-front
 transitions.

Like the Ginzburg-Landau equations with intrinsic gain,
the parametrically
driven equation (\ref{eq:ddcgle_bis}) 
arises in a wide range
 of physical applications. These include nonlinear Faraday resonance
 in  vertically vibrated layers of water \cite{Miles,Faraday,Elphick_Meron}
 and nonlinear lattices \cite{lattices};
 commensurate-incommensurate transitions
 in convective systems \cite{Coullet_1986};
 waves in nematic and cholesteric liquid crystals in rotating 
 magnetic fields \cite{liquid_crystals};
 magnetisation waves in easy-plane ferromagnet exposed to microwave
fields \cite{BBK};  domain walls in the easy-axis ferromagnet near
the Curie temperature \cite{Curie} and in the easy-plane magnet in the
stationary  magnetic field \cite{BWZ};
 nonlinear fiber lines with phase-sensitive
 amplification and
 mean-field models of degenerate optical parametric 
 oscillators under continuous-wave pumping
 \cite{Optics_NLS};
 pulsed optical parametric oscillators with spectral
 filtering and lasers with intracavity parametric amplification 
 \cite{Longhi_GL}.
In  most cases the models considered in literature 
are either purely diffusive
($c=\infty$)
\cite{Longhi_GL,Curie,Coullet_1986,liquid_crystals} 
or purely dispersive ($c=0$)
\cite{Miles,Faraday,Elphick_Meron,Optics_NLS,BBK,BWZ}.
However there are situations where
 it is crucial that $c$ be finite but nonzero. One
 physical phenomenon to which both diffusion and dispersion 
 make essential contributions, is Faraday resonance in strongly
 dissipative media such as viscous fluids and granular materials
 \cite{Tsimring}. Another situation  where $0<c< \infty$, corresponds to
  nondegenerate optical parametric oscillators
 \cite{Longhi_nondegenerate} and  fiber-optic telecommunication
links; in these contexts, the term $-ic\psi _{xx}$
represents spectral filtering in the spatial or temporal domain,
respectively. In fact the  applicability of
 Eq.(\ref{eq:ddcgle_bis}) with nonzero $c$ and $g$ may happen to be even wider;
it is commonly held that this equation
 provides a phenomenological 
description to a broad range of pattern-forming systems
\cite{nonzero_c}.

To study the solitary wave phenomenology introduced by taking  the
($-ic\psi_{xx}$)-term
into account, we will use the diffusion-free limit ($c=0$) as
a starting point, and perform the numerical continuation of 
analytical solutions available in that case, to nonzero $c$.
The stability of solutions obtained in this way will also be studied 
numerically. We will show that ``switching on" the diffusion gives 
rise to a sequence of stable multihumped
pulses occurring in the vicinity of a certain particular value
of the diffusion coefficient,  $c_{\rm lim}=c_{\rm lim}(h,\gamma)$.
The closer  $c$ is to $c_{\rm lim}$, the greater is the number of
multihump solutions coexisting at this $c$. 
 The
 solutions with more than 5 or 6 humps 
 describe a flat ``plateau"
(where $\psi$ is equal to the stable  flat nonzero solution)
 sandwiched between two fronts of
opposite polarity.

The paper is organised as follows. Section  \ref{flat_backgrounds}
deals mainly with spatially homogeneous solutions. In particular, we
show that there is a stable flat nonzero solution for sufficiently large
$c$, and this uniform solution can serve as a background for
solitary waves.
In section \ref{continua}
we use perturbation-type arguments to demonstrate that both $\psi_+$
and $\psi_-$ solitons are continuable in $c$ and in section
\ref{sec:adiabatic} we construct the 
 solutions with nonzero $c$ in the adiabatic approximation.
The actual continuation is carried out numerically in section
\ref{sec:continuation} where we also examine
the  stability  of the continued
solutions. Some additional
insight into the bifurcation of stationary pulses
is gained in section \ref{h_vs_c}.
Finally,
section \ref{sec:conclusion} summarises results of this work.

We close this introduction by mentioning a recent 
 publication \cite{Sakaguchi} which was devoted to
the study of the single-humped solution 
 of Eq.(\ref{eq:ddcgle_bis}),
 by means of an averaging technique and direct
numerical simulations. Neither 
the multistability 
of  pulses  nor the pulse-front transitions were 
dealt with in Ref.\cite{Sakaguchi}.

\section{Existence and stability inequalities}
\label{flat_backgrounds}

The transformation $\psi(x,t) \to e^{-i \omega t} \psi(x,t)$
casts Eq.(\ref{eq:ddcgle_bis})
in a `standard' autonomous form
\begin{equation}
i\psi _{t}
+(1-ic)\psi _{xx}+ 2\left| \psi \right|^{2}\psi -\psi
=h\psi ^{\ast }-i\gamma \psi,
\label{eq:ddcgle}
\end{equation}
where we set $g=0$ and rescaled $t$ so that $\omega =1$. This is 
the representation that we are going to work with in what follows.

The  equation (\ref{eq:ddcgle}) has three
time-independent spatially uniform
  solutions, or `flat backgrounds', for short.
(We do not distinguish between solutions
different only in the overall sign here.) One flat solution is
 $\psi _{{\rm flat}}=0$; it will be central for this
  work  where we are focussing on solutions decaying to zero
at infinities.
  The other two flat solutions are given by
\begin{equation}
\psi _{{\rm flat}}=\Psi_{\pm}^{(0)} \equiv
\left( A_{\pm }/\sqrt{2}\right) e^{-i\Theta _{\pm
}} \,,
\label{flat_back}
\end{equation}
where 
\begin{eqnarray}
A_{\pm }=\sqrt{1\pm \sqrt{h^{2}-\gamma ^{2}}},
\quad
 2\Theta _{\pm }= \arccos{(\pm
\sqrt{1-\gamma ^{2}/h^{2}})}.
\label{plusminus}
\end{eqnarray}
Note that these solutions do not depend on $c$.

\subsection{Stability of spatially uniform solutions}
\label{spatially_uniform}
  One can easily check
  that the zero solution
  is stable as long as $h < \sqrt{1+\gamma^2}$, irrespectively
  of the value of $c \ge 0$. The analysis of the flat nonzero
  solutions is somewhat more laborious.

Linearising
 Eq.(\ref{eq:ddcgle}) about $\psi _{{\rm flat}}=\Psi_{\pm}^{(0)}$,
 and assuming a perturbation
$\delta \psi  \propto \exp \left[ i(\omega
t-kx)\right] $ yields the dispersion relation
\begin{equation}
i\omega =-(ck^{2}+\gamma )\pm i\sqrt{Z}\;,
\label{eq:dispersioneqn}
\end{equation}
where
\begin{equation}
Z = \left( 1-2A^{2}+k^{2}\right) ^{2}+A^{2}\left( A^{2}-2\right)
-h^{2}\;.  \label{eq:Zeqn}
\end{equation}
(Here $A$ stands for $A_+$ or $A_-$,
depending on which solution  we are
linearizing about.)  The flat solution is stable iff
${\rm Re}(i\omega )\leq 0$,
i.e., when  $Z\geq -(ck^{2}+\gamma )^{2}$, for all real $k$.
The latter condition amounts to an inequality
\begin{equation}
\left( 1+c^{2}\right) s^{2}+2\left( 1-2A^{2}
+\gamma c\right) s+4A^{2}\left(
A^{2}-1\right) \geq 0,
\label{inequalityfork}
\end{equation}
where $s$ stands for $k^2$.

Let, first, $\psi=\Psi_-^{(0)}$.
Since $A_-^{2}<1$, the inequality (\ref{inequalityfork})
does not hold for $k=0$; hence 
the `low' background $\Psi_-^{(0)}$ is always unstable.

Let now $\psi=
\Psi_+^{(0)}$, the `high' background. The quadratic
expression in (\ref{inequalityfork}) will be positive
for all $s \ge 0$ if either its two roots $s_1$ and $s_2$ are both real
negative or complex. (Note that we cannot have two real roots
of opposite signs as the quadratic's constant term,
$4A^{2}\left(A^{2}-1\right)$, is always positive for $\psi=\Psi_+^{(0)}$.)
Whether the roots are real or complex is determined by
the discriminant of the quadratic (\ref{inequalityfork})
which can be written as
\[
{\cal D}% = [\gamma^2 -4A_+^2(A_+^2-1)]c^2 +2(1-2A_+^2) \gamma c+1
=[\gamma^2 -4A_+^2(A_+^2-1)](c-c_+)(c-c_-). 
\]
Here we have introduced
\begin{equation}
c_\pm=
\frac{\left( 2A_{+}^{2}-1\right) \gamma  \pm
2A_+ \sqrt{\left(
A_{+}^{2}-1\right) \left( 1+\gamma ^{2}\right) }}
{\gamma^{2}-4A_{+}^{2}\left( A_{+}^{2}-1\right) }.
\label{eq:cpmroots}
\end{equation}

We need to consider two cases.
Assume, first, that $4A_+^2(A_+^2-1)< \gamma^2$. In this case
the discriminant  is
negative (and hence, the roots $s_{1,2}$ are complex) provided
$c$ lies in the interval $c_-<c<c_+$.
On the other hand, the roots $s_{1,2}$ are real negative
in this case if ${\cal D}$ is $\geq 0$ and the 
coefficient in front of the middle term in
(\ref{inequalityfork}) is positive:
\begin{equation}
c > c_0 \equiv \left( 2A_{+}^{2}-1\right) /\gamma.
\label{c0}
\end{equation}
Since in the case at hand we have $0<c_-<c_0<c_+$, the union of the
above two
`stable' regions, $c_-<c<c_+$ and $c>c_+$, with the
endpoints included, is simply  $c \ge c_-$.

The second case is defined by inequality
$4A_+^2(A_+^2-1)> \gamma^2$. Here we have  $c_+<0<c_-<c_0$
and
the quadratic (\ref{inequalityfork}) cannot have negative real roots
as the region  (\ref{c0}) does not overlap with the region where
${\cal D} \geq 0$.
However, it can have complex roots --- provided $c>c_-$.

Thus we arrive at a
simple stability criterion for the flat nonzero solution $\Psi_+^{(0)}$,
valid for all $h$ and $\gamma$:
\begin{equation}
c \geq c_-(h, \gamma),
\label{flat_stab}
\end{equation}
with $c_-$ as in (\ref{eq:cpmroots}). Note that the stability
threshold $c_-(h, \gamma)$ is always strictly positive.  This
implies, in
particular,  that the solution $\Psi_+^{(0)}$ is
always  unstable in the case $c=0$.
(This fact has already been mentioned in the literature
\cite{LonghiGeraci}.)
Therefore the `focusing', or `attractive',
damped-driven Schr\"odinger equation (Eq.(\ref{eq:ddcgle}) with $c=0$)
does not have stable flat backgrounds except
the trivial one, $\psi=0$. The analysis of 
localised solutions over flat {\it
nonzero\/}
backgrounds becomes meaningful only within 
the full Ginzburg-Landau equation, i.e. Eq.(\ref{eq:ddcgle}) with
positive $c$.

\subsection{The flat solutions as backgrounds to solitary waves}

No less important is the
question of when  a flat solution can serve as an asymptotic value
to a
localised waveform --- in other words, when is  spatial
decay to the flat solution possible.
To find  the corresponding criterion, we
set $\omega =0$ in equation (\ref{eq:dispersioneqn}). This results in a
quadratic equation
\begin{equation}
\left( 1+c^{2}\right) s^{2}-2\left( 2A^{2}-1-\gamma c\right)
s+4A^{2}\left( A^{2}-1\right) =0,
\label{eq:squad}
\end{equation}
where $s=k^2$. The spatial decay to a flat
 background  is possible unless  both roots of Eq.(\ref{eq:squad}),
 $s_1$ and $s_2$, are positive.

 In the case of the  $\Psi_-^{(0)}$ background,
 the discriminant of equation
 (\ref{eq:squad}) is positive while
 the constant term  is negative, whence
 $s_1>0$ and $s_2<0$. Consequently, the decay to $\Psi_-^{(0)}$
 may occur for
all values of $h$, $\gamma $ and $c$.
(This fact is of little importance, however, 
since we have just shown that this background is
always unstable.)

In the case of the  $\Psi_+^{(0)}$ solution, we {\it may\/}
have two positive
roots --- provided ${\cal D}>0$ and $c<c_0$, with $c_0$ as in
(\ref{c0}). Following the steps in the previous
subsection, one can readily show that this situation
occurs only if $0<c<c_-$, with $c_-$ as in
(\ref{eq:cpmroots}). Thus the {\it stable\/} constant
solutions,
defined by inequality (\ref{flat_stab}), can always serve
as  backgrounds to  fronts and pulses.
We will come across localised solutions over nonvanishing
backgrounds in section \ref{sec:continuation}  below.

Finally, to examine the case of
the zero background, we set both $A=0$ and $\omega =0$ in
equation (\ref{eq:dispersioneqn}). This
yields
\begin{equation}
(1+c^2) s^2+2 (\gamma c+1)s+ \gamma^2-h^2=0,
\label{zero_decay}
\end{equation}
where, as before, $s=k^2$. Since the middle term in  (\ref{zero_decay})
has a positive coefficient, we have $s_1+s_2<0$,
which means that the situation where both $s_1$ and $s_2$ are
positive, can never occur in this case.
Thus the decay to the
zero  background is possible for all $h$, $\gamma$ and $c$.

\subsection{The threshold driving strength for pulses}
\label{sec:threshold}

Our last  result in this section concerns
the range of driving strengths which can support 
localised solutions in the presence of dissipation.
Multiplying Eq.(\ref{eq:ddcgle}) by $\psi ^{\ast }$ and
subtracting the complex conjugate of the resulting equation
gives
what would be a local conservation law of the number of
particles if $c$, $h$ and $\gamma$ were equal to zero:
\begin{eqnarray}
i\partial_t |\psi|^2 +  \left( \psi
_{x}\psi ^{\ast }-\psi _{x}^{\ast }\psi \right)_x
\nonumber \\
=h\left[ \left( \psi ^{\ast }\right)
^{2}-\psi ^{2}\right]
-2i\gamma \left| \psi \right| ^{2}
+ic\left( \psi _{xx}\psi
^{\ast }+\psi _{xx}^{\ast }\psi \right) \,.
\label{eq:ada-eq1x}
\end{eqnarray}
Assuming that $\psi$ and $\psi_x \to 0$ as $|x| \to \infty$
and integrating  (\ref{eq:ada-eq1x}) over the real line, we get
\begin{equation}
\frac{d}{dt} \int |\psi|^2 dx=
2 \int |\psi|^2 [h \sin(2 \chi)- \gamma] dx-2c \int|\psi_x|^2 dx,
\label{decay}
\end{equation}
where  we have denoted $-\chi(x,t)$  the phase of the
complex field $\psi$: $\psi=|\psi| e^{-i \chi}$.
Let  $h < \gamma$ (and remember that $c>0$).
In this case it follows from (\ref{decay}) that the time derivative of
$ \int |\psi|^2dx $ remains negative at all times. Hence as
$t \to \infty$,
$\psi(x,t) \to 0$ for all $x$. No stationary,  
time-periodic, quasiperiodic  or chaotic solutions, decaying to
zero as $|x| \to \infty$, can
arise if $h < \gamma$. We will make use of this simple criterion in
what follows.

\section{Continuability of the two NLS solitons}
\label{continua}

In the limit $c=0$, Eq.(\ref{eq:ddcgle}) becomes the parametrically driven
damped nonlinear Schr\"{o}dinger (NLS) equation, which has  exact
time-independent solitary wave solutions \cite{Miles,BBK} of the form
\begin{equation}
\psi _{\pm }(x,t)=A_{\pm }
 \, e^{-i\Theta_{\pm }} \,
{\rm sech}\left( A_{\pm }x\right),
\label{eq:psipm}
\end{equation}
with $A_{\pm}$ and $\Theta_{\pm}$ as in (\ref{plusminus}).
The solution $\psi _{-}$ is unstable for all $h$ and $\gamma $,
while $\psi _{+}$ is stable in a certain parameter region
\cite{BBK}.

The purpose of
 this section is to show that the
 solitary pulse solutions $\psi _{\pm }$
of the NLS equation   persist
 for $c\neq 0$. We restrict ourselves to stationary
solutions ($\psi _{t}=0$).
Writing $\psi =\phi e^{-i\Theta} $ with $\Theta $  a constant phase to be
chosen later, Eq.(\ref{eq:ddcgle}) becomes an 
ordinary differential equation for $\phi$:
\begin{equation}
\left( 1-ic\right) \phi _{xx}+2\left| \phi \right| ^{2}\phi -\left(
1-i\gamma \right) \phi =h\phi ^{\ast }e^{2i\Theta}.
\label{eq:eqnforphi}
\end{equation}
To find out whether  solutions available
at $c=0$ can be continued to nonzero $c$,
we expand $\phi $ in power series
$\phi =\phi _{0}+c\phi _{1}+c^{2}\phi _{2}+\ldots$,
substitute into Eq.(\ref{eq:eqnforphi}) and match like
powers of $c$. It is
convenient to choose $\Theta $ to
be $\Theta _{+}$ in the case of the soliton $\psi _{+}$, 
and $\Theta _{-}$ for $\psi _{-}$;
this choice
makes $\phi _{0}$ real. Matching terms
linear in $c$
and decomposing $\phi _{1}$ into its real and
imaginary part ($\phi _{1}=u+iv$),
yields an equation
\begin{equation}
L_{\pm }\left(
\begin{array}{c}
u \\
v
\end{array}
\right) =\left(
\begin{array}{c}
0 \\
-\phi _{0}^{\prime \prime }
\end{array}
\right) \,,
\label{eq:eqnUV}
\end{equation}
where
 the primes stand for the derivatives in $x$ and
 the  operators $L_{\pm}$ are defined by
\begin{widetext} \begin{equation}
L_{\pm }=\left(
\begin{array}{cc}
-\partial _{x}^{2}+1-6\phi _{0}^{2}+h\cos \left( 2\Theta _{\pm }\right) &
2\gamma \\
0 & -\partial _{x}^{2}+1-2\phi _{0}^{2}-h\cos \left( 2\Theta _{\pm }\right)
\end{array}
\right).
 \label{eq:sim-eqnforLpm}
\end{equation}
 \end{widetext}(The subscripts $+$ and $-$ indicate whether we are checking the
continuability of $\psi _{+}$ or $\psi _{-}$.)

According to  Fredholm's alternative, Eq.(\ref{eq:eqnUV})
has bounded solutions if and only if its right-hand side is orthogonal to
the kernel of the adjoint operator $L_{\pm }^{\dagger }$.
Transforming  to $\xi \equiv
A_+x$ and $\xi \equiv
A_-x$ in the case of the $\psi _{+}$
and $\psi _{-}$  solution, respectively, the equation for the zero modes
spanning  $\mbox{ker} \, L_{\pm }^{\dagger }$ becomes
\begin{equation}
\left(
\begin{array}{cc}
L_{1} & 0 \\
2\gamma & L_{0}-\epsilon _{\pm }
\end{array}
\right) \left(
\begin{array}{c}
f \\
g
\end{array}
\right) =0\,,
\label{eq:ker-l-dag}
\end{equation}
where $f=f(\xi)$, $g=g(\xi)$,
\[
\epsilon _{\pm } \equiv 2\frac{A_{\pm }^{2}-1}{A_{\pm }^{2}}=\pm \,2\,\frac{%
\sqrt{h^{2}-\gamma ^{2}}}{1\pm \sqrt{h^{2}-\gamma ^{2}}},
\]
and we have introduced the standard P\"oschl-Teller
operators with familiar spectral properties:
\begin{equation}
L_{0}= -\partial _{\xi }^{2}+1-2{\rm sech}^{2}\xi,
\quad
L_{1}=-\partial _{\xi }^{2}+1-6{\rm sech}^{2}\xi.
\nonumber
\end{equation}

The only discrete eigenvalue of $L_{0}$
 is $E_0=0$, while
the continuous spectrum  occupies the
semiaxis $E \ge1$.
Therefore, the operator
$\left( L_{0}-\epsilon_{\pm } \right) $ is invertible
as long as $\epsilon _{\pm }<1$ and $\epsilon _{\pm }\neq 0$.
Assuming that the zero background is stable
(i.e., assuming that  $h^{2}<1+\gamma ^{2}$), the quantity $\epsilon_{+}$
is indeed less than one but
 greater than zero.
 On the other hand, the $\epsilon_{-}$ is negative.  (An exception
is the point $h=\gamma $, where we have $\epsilon _{\pm }=0$.)
 Thus we conclude that $\left( L_{0}-\epsilon _{\pm}\right)$
is invertible except in the special case
 $h=\gamma $.
 (However, even in this
special case the operator $L_0$ is invertible on
{\it odd\/} functions
because the
eigenfunction  ${\rm sech\/} \, \xi $ associated
with the zero eigenvalue is even.)
  Consequently, $f(\xi)$ cannot be equal to zero for $h \neq \gamma$
--- otherwise the bottom equation in (\ref{eq:ker-l-dag})
would imply that $g(\xi)=0$, too.
Fortunately,
the operator $L_{1}$ does have a zero eigenvalue,  and
therefore $f(\xi)$ can be a nonzero multiple of the corresponding
 eigenfunction (which is $\tanh \xi \, {\rm sech} \, \xi $.)

Thus in the case $h \neq \gamma $ the kernel of
$L_{\pm}^{\dagger }$ is spanned by just one zero mode, namely
\begin{equation}
\left(
\begin{array}{c}
f_1 \\
g_1
\end{array}
\right) = \left(
\begin{array}{c}
\tanh \xi \, {\rm sech} \, \xi \\
-2\,\gamma \left( L_{0}-\epsilon _{\pm }\right) ^{-1}\,\left( \tanh \xi
\, {\rm sech}\, \xi \right)
\end{array}
\right) \,.
\label{eq:zerom-hneqg}
\end{equation}
On the other hand, when $h=\gamma $, the dimension of
the kernel space is two. Firstly, the zero mode (\ref{eq:zerom-hneqg})
persists for $\epsilon_{\pm}=0$ as the operator $L_0^{-1}$
is defined on odd functions.
Secondly, there is another
 zero mode  given by
\begin{equation}
\left(
\begin{array}{c}
f_2 \\
g_2
\end{array}
\right) =\left(
\begin{array}{c}
0 \\
{\rm sech} \, \xi
\end{array}
\right)
 \,.
\label{eq:zerom}
\end{equation}

It is obvious
 that the vector-function
 (\ref{eq:zerom-hneqg}), whose  both components are
odd functions of $\xi$, is orthogonal to
the right-hand side of Eq.(\ref{eq:eqnUV}) ---
which is even.
Hence Eq.(\ref{eq:eqnUV}) is solvable for
$h\neq \gamma $.
 It is also easy to check that the mode
(\ref{eq:zerom}) is {\it not\/} orthogonal to
the right-hand side of (\ref{eq:eqnUV})
and so the solvability condition is {\it not\/} satisfied for
 $h=\gamma $.

Thus, having
started from the two nonlinear Schr\"odinger solitons,
 we constructed two one-pulse solutions of the Ginzburg-Landau
equation
to within the first order in the small parameter $c$:
$\psi=\psi_\pm + c e^{-i \Theta_\pm} (u+iv)+...$.
Consequently, we expect to be able to continue the Schr\"odinger
solitons $\psi_{\pm }$
into the region $c \neq 0$ (provided that $h\neq \gamma $.)
This expectation
is born out by results displayed  in section \ref{sec:continuation}
below,
which present an outcome of the {\it numerical\/} continuation
of $\psi _{\pm }$ in $c$.

\section{Adiabatic Approximation}
\label{sec:adiabatic}

Before attempting the full-scale numerical continuation, it is
instructive to construct approximate solutions. Our
approximation will be valid for small $c$ and 
 exploit the fact that
when $c=0$, the stationary pulse-like
solutions of equation (\ref{eq:ddcgle})
have the form
\begin{equation}
\psi =a\,{\rm sech}\left( ax\right) e^{-i\theta},
\label{eq:ada-soln}
\end{equation}
where $a=A_\pm$ and $\theta=\Theta_\pm $ are constants
defined by Eqs.(\ref{plusminus}). For $c$ small nonzero,
approximate
solutions can be sought for by assuming that $\psi$ retains 
the form (\ref{eq:ada-soln}),
but $a$ and $\theta $ become slowly varying functions of $t$.

To obtain an expression for $\dot{a}$ (the overdot stands for the derivative
in $t$), we substitute the ansatz
 (\ref{eq:ada-soln}) into (\ref{eq:ada-eq1x}). Integrating
over $x$ and using the boundary conditions $\psi _{x}\rightarrow 0$ as $%
|x|\rightarrow \infty $ produces an evolution equation for  the
pulse's amplitude:
\begin{equation}
\dot{a}=2a(h\sin 2\theta -\gamma -\tilde{c}a^{2}) \, ,
%\dot{a}=2ha\sin (2\theta) -2\gamma a-2\tilde{c}a^{3} \, ,
\label{eq:ada-eq1}
\end{equation}
where  $\tilde{c} \equiv c/3$.
An equation for the
pulse's phase arises by multiplying (\ref{eq:ddcgle})
by $\psi^{\ast}$,
adding with its complex conjugate,
substituting (\ref{eq:ada-soln}) for $\psi$
and  integrating over $x$:
\begin{equation}
\dot{\theta}=h\cos 2\theta +1-a^{2}\,.
 \label{eq:ada-eq2}
\end{equation}

Fixed points of the system (\ref{eq:ada-eq1}) and (\ref{eq:ada-eq2})
correspond to
 stationary solutions of 
(\ref{eq:ddcgle}).
These can be easily found 
explicitly:
Eliminating $\theta$ from the stationary system
\begin{equation}
h \sin (2\theta) =\tilde{c}a^{2}+\gamma, \quad
h\cos (2\theta) =a^{2}-1,  \label{eq:stat-eq2}
\end{equation}
produces a quadratic equation
\begin{equation}
\left( 1+\tilde{c}^{2}\right) a^{4}+2\left( \gamma \tilde{c}-1\right)
a^{2}+ 1+\gamma ^{2}-h^{2} =0
\label{eq:stat-eqnforA}
\end{equation}
with roots
\begin{equation}
a_\pm^{2}=\frac{\left( 1-\gamma \tilde{c}\right) \pm \sqrt{
( h^2-1) \tilde{c}^2 -2\gamma \tilde{c}
+ h^{2}-\gamma^2 
}}
{1+\tilde{c}^{2}}\,.
\label{eq:ada-soln-for-A}
\end{equation}
The corresponding $\theta_\pm$ are defined by their sine and
cosine in (\ref{eq:stat-eq2}).

It is not difficult to determine when the roots 
(\ref{eq:ada-soln-for-A}) are real
and positive.
We assume
that the zero background solution of Eq.(\ref{eq:ddcgle}) is stable, i.e.
the constant term in (\ref{eq:stat-eqnforA}) is positive.
Hence if the roots are real, they are of the same sign,
and this sign is opposite to that of the middle term in
(\ref{eq:stat-eqnforA}). Therefore we have two positive roots
provided the discriminant
\begin{equation}
{\cal D} =\left( h^{2}-1 \right) \tilde{c}^{2}-2\gamma \tilde{c}
+ h^{2}-\gamma^{2}  \equiv
\left( h^{2}-1\right) (\tilde{c}-\tilde{c_1})( \tilde{c} -   \tilde{c_2})
\label{eq:ada-realAA}
\end{equation}
is nonnegative, and, at the same time, the inequality
$\gamma \tilde{c} <1$ holds true.  In (\ref{eq:ada-realAA})
we have introduced
\begin{equation}
\tilde{c}_{1,2}=\frac{\gamma \pm h\sqrt{ 1+\gamma
^{2}-h^{2} }}{h^{2}-1}\,,
\label{eq:ada-soln-for-c}
\end{equation}
where the  $+$ corresponds to ${\tilde c_1}$ and $-$ to ${\tilde
c_2}$.

It is straightforward to verify that for small $h$, $h^2<1$,
we have $\tilde{c_1}< 0<\tilde{c_2}< \frac{1}{\gamma}$,
while for larger $h$,  $h^2>1$,
we have $0< \tilde{c_2} < \frac{1}{\gamma} < \tilde{c_1} $.
(Here we have made use of the inequality $h> \gamma$;
as proved in section \ref{sec:threshold}, no stationary localised
solutions exist for $h< \gamma$.)
Therefore, the region of
$c$ values where the roots of (\ref{eq:stat-eqnforA}) are
positive, is given by the
inequality $\tilde{c}<\tilde{c_2}(h, \gamma)$ --- for all $h$.

Thus we  conclude that the adiabatic
equations have two stationary points
for $c <3\tilde{c_2}(h, \gamma)$, and none for $c >3\tilde{c_2}$.
We complete the adiabatic analysis by classifying their
stability and bifurcation.

Linearising equations
(\ref{eq:ada-eq2}) about the stationary points
and assuming  small perturbations of the form $\delta a
= C_1 e^{2 \lambda t}$ and $\delta
\theta =C_2e^{2 \lambda t}$, yields  a characteristic
equation
\[
\lambda^2 +(\gamma+c a^2) \lambda +
2ha^2 \cos (2 \theta) -h \sin (2 \theta) [h \sin (2 \theta) -
\gamma -ca^2]=0.
\]
Since the coefficient in front of the middle term 
in this equation is positive,
either its two roots are complex with  negative
real parts, or we have two real roots, of which one is
negative. The case when the second root is positive
(and hence the fixed point is unstable) occurs if the
constant term is negative. Conversely, if the
constant term is nonnegative:
\begin{equation}
2ha^2 \cos (2 \theta) -h \sin (2 \theta) [h \sin (2 \theta) -
\gamma -ca^2] \ge 0,
\label{stabby}
\end{equation}
 the fixed point is
stable. Simplifying Eq.(\ref{stabby}), we get the stability condition
in the form
\[
a^{2} \geq \frac{1-\gamma \tilde{c}}{1+\tilde{c}^{2}}\,.
\]
Comparing this to the expressions for
 $a_{\pm }$ in (\ref{eq:ada-soln-for-A}),
we conclude that
 for all $c$, $h$ and $\gamma$,
the  points $(a_{+},\theta_+)$  and $(a_{-},\theta_-)$
are a stable node and saddle, respectively.

The two fixed points come into being through a saddle-node
bifurcation which occurs 
as the diffusion coefficient $c$ is decreased past $c=3{\tilde c}_2$
for the fixed $h$ and $\gamma$ or, alternatively, as
 the driving amplitude
 $h$ is increased for the fixed dissipation coefficients
$c$ and $\gamma$. 
One can easily find the bifurcation value of $h$, at which 
two complex roots of the quadratic equation (\ref{eq:stat-eqnforA})
converge on the positive real axis. (Here we are assuming that ${\tilde c}$
is smaller than $\frac{1}{\gamma}$.) Writing the discriminant 
(\ref{eq:ada-realAA}) as
\[
{\cal D}= ({\tilde c}^2+1) 
\left[ h^2 -\frac{({\tilde c}+\gamma)^2}{{\tilde c}^2+1}
\right],
\]
the threshold driving strength is given by
\begin{equation}
h_{\rm ad}= \frac{{\tilde c}+ \gamma}{\sqrt{{\tilde c}^2+1}},
\label{threshold}
\end{equation}
where the subscript ``ad" serves to remind that Eq.(\ref{threshold})
was obtained in  the adiabatic approximation.
It is important to emphasise that this formula is valid
only for small ${\tilde c}$. Equivalently, the expression 
(\ref{eq:ada-soln-for-c}) for the turning point ${\tilde c}_2$ 
is valid only for $h$ close to $\gamma$.

\section{Numerical continuation and stability analysis}
\label{sec:continuation}

\begin{figure}
\includegraphics[ height = 2in, width = 0.5\linewidth]{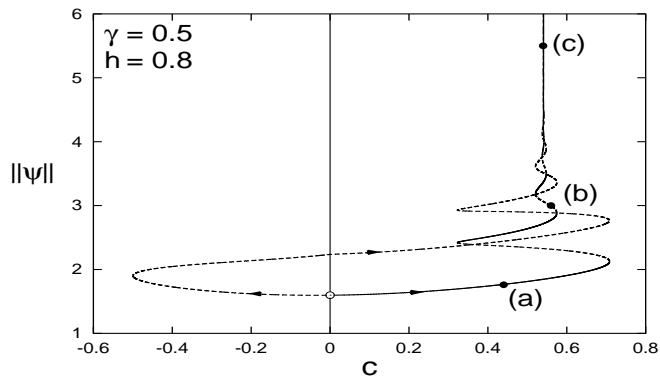}
\caption{\sf The bifurcation diagram displaying the Sobolev
norm versus the diffusion 
coefficient,  for the solitary
pulse obtained by the continuation of the 
stable soliton $\psi_+$ in $c$. The continuation 
starts from a point on the $(c=0)$-axis, marked by an open
circle.  The arrows indicate the 
directions of continuation
and are only added for reference purposes.  Solutions at points marked by the black dots (a), (b)
and (c) are 
shown in Figure \ref{conti}. The solid lines correspond to
stable and dashed curves to unstable branches.
 }
\label{biffy}
\end{figure}

In this section we describe the 
bifurcation diagram obtained by the numerical continuation 
of the solitons $\psi_\pm$ in the parameter $c$. The diagram is
presented in Fig.\ref{biffy} and displays the Sobolev norm of
the solution, 
\[
||\psi|| = \sqrt{ \int (|\psi_x|^2 + |\psi|^2) dx}
\]
as a function of $c$.

\subsection{The method}
For stationary solutions, $\psi=\psi(x)$, equation (\ref{eq:ddcgle})
reduces to an ordinary differential equation
\begin{equation}
(1-ic)\psi _{xx}+2\left| \psi \right| ^{2}\psi -(1-i\gamma )\psi
=h\psi ^{\ast }.
\label{stat_GLE}
\end{equation}
 For
the numerical continuation of solutions to Eq.(\ref{stat_GLE}) 
we utilised   
the AUTO97 software package \cite{AUTO97}. The infinite line
was approximated by a finite interval $(-L,L)$, with $L=100$,
and the boundary conditions $\psi(\pm L)=0$. The tolerance 
of the computation (the grid norm of the 
difference between the right and left-hand side 
of (\ref{stat_GLE})) was set to be $10^{-7}$.

The solid and broken curves in Fig.\ref{biffy} represent
stable and unstable branches,
respectively. The stability  was examined
 by linearising Eq.(\ref{eq:ddcgle}) about the corresponding
stationary solution.
Choosing the small perturbation in the form
$\delta \psi (x,t)=\left[  u(x)+i v(x)\right] e^{\lambda t}$,
with real $ u$ and $ v$, we 
arrive at an eigenvalue problem
\begin{equation}
{\cal H}\left(
\begin{array}{c}
 u \\
 v
\end{array}
\right) =\lambda J\left(
\begin{array}{c}
 u \\
 v
\end{array}
\right) \;{\rm ,}
\label{EV_problem}
\end{equation}
where 
\begin{widetext}
\[
{\cal H}=
\left(
\begin{array}{lr}
- \partial_x^2+1+h-6{\cal R}^2-2{\cal I}^2
&
-c\partial_x^2+\gamma -4{\cal R} {\cal I}
\\
c\partial_x^2-\gamma -4 {\cal R} {\cal I}
&
-\partial_x^2+1-h-2{\cal R}^2-6{\cal I}^2
\end{array}
\right),
\]
\end{widetext}
and
\[
J=\left(
\begin{array}{lr}
0 & -1 \\
1 & 0
\end{array}
\right).
\]
The $\cal R$ and $\cal I$ are the real and imaginary parts
of the solution: $\psi={\cal R} +i{\cal I}$.
We solved the eigenvalue problem by expanding $u$ and  $v$ over
500
Fourier modes in the interval $(-100,100)$.

\begin{figure}
\includegraphics[ height = 2in, width = 0.5\linewidth]{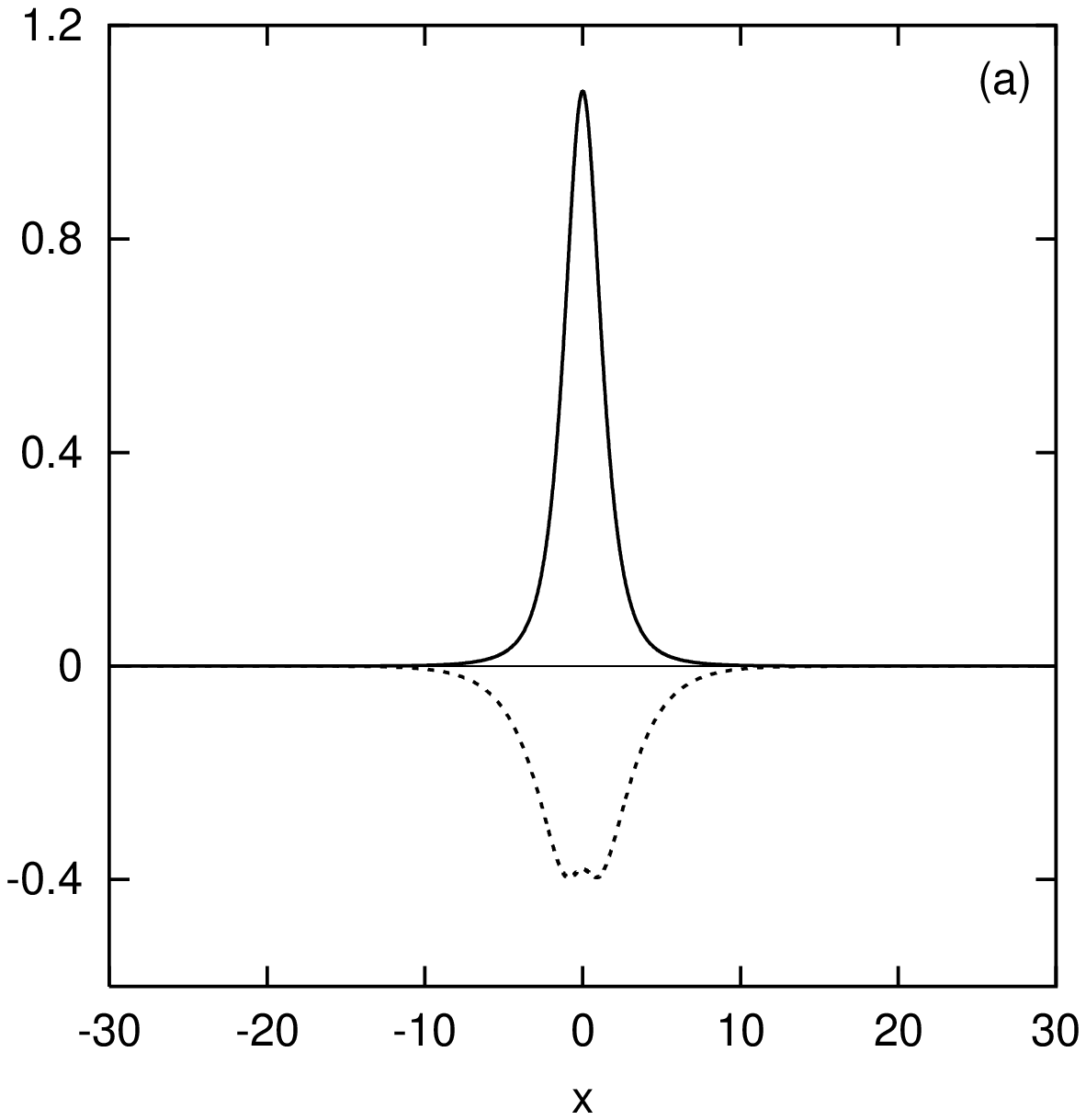}
\includegraphics[ height = 2in, width = 0.5\linewidth]{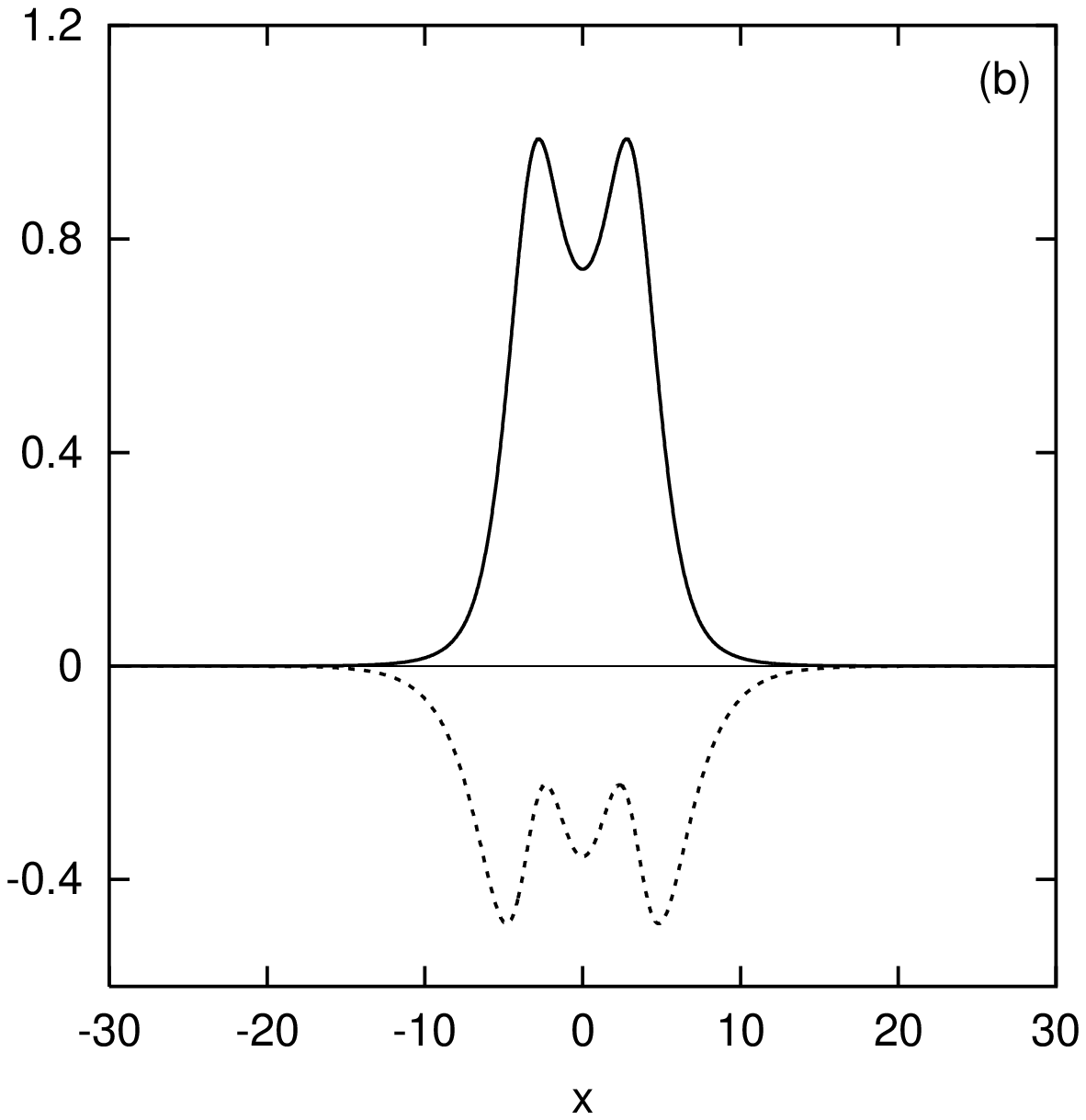}
\includegraphics[ height = 2in, width = 0.5\linewidth]{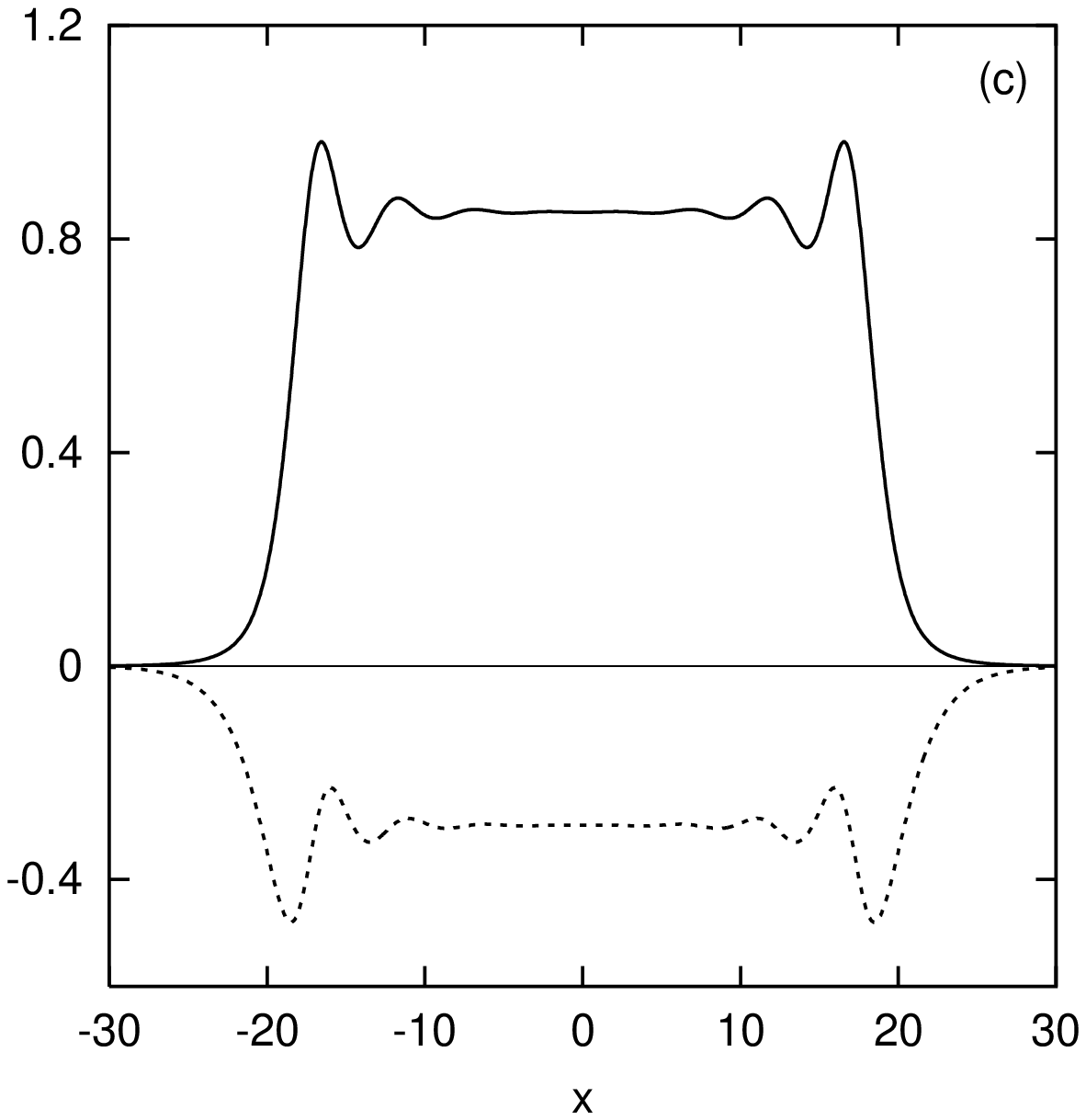}
\caption{\sf (a),(b),(c): Solutions at the 
corresponding points in
Fig.\ref{biffy}. Solid line: real part; dashed line: imaginary part.
 }
\label{conti}
\end{figure}
%%%%%%%%%%%%%%%%%%%%%%%%%%%%%%%%%%%%%%%%%%%%%%%%%%%%%%%%%%%%%%%%%%%%%%%%%

\subsection{Continuation and stability}
 The continuation in $c$ was
performed for fixed values of $h$ and $\gamma$. We selected
$\gamma =0.5$ and $h=0.8$; for these $h$ and $\gamma$ the nonlinear
Schr\"odinger soliton $\psi_+$  is stable \cite{BBK}.
Before proceeding to the $\psi_+$
soliton, however,
 we briefly deal with the $\psi_-$-case.
In agreement with predictions of section
\ref{continua}, the soliton $\psi_-$  was found to persist 
both for $c<0$ and $c>0$.  We were in fact able to continue it
indefinitely without encountering any obstacles
in both directions. 
 As $ c \to \infty$, the width of the localised solution 
of (\ref{stat_GLE}) grows and the solution
tends to $\phi_{\gamma,h}({x}/\sqrt{c})$,
where  
$\phi_{\gamma,h}(X)$ stands for a pulse-shaped solution of equation
\begin{equation}
 - i\phi_{XX}+2\left| \phi \right| ^{2}\phi -(1-i\gamma )\phi
=h\phi ^{\ast }.
\label{c_infty}
\end{equation}
In a similar way, as $ c \to -\infty$, the solution
tends to $\phi^*_{-\gamma,h}({x}/\sqrt{-c})$.
When 
continued to $c>0$, the  pulse remained 
unstable for all $c$, 
with a single positive eigenvalue in its spectrum.
(As for the negative-$c$ region, all solutions there 
are {\it a priori\/} unstable against continuous spectrum
perturbations with arbitrarily large ${\rm Re\/} \lambda$.)

The continuation of the $\psi_+$ soliton
proved to be more rewarding from the stability viewpoint.
The corresponding bifurcation diagram is
displayed in Fig.\ref{biffy}. As with the  soliton $\psi_-$, the
$\psi_+$ persists both for $c > 0$ and $c < 0$ --- in agreement 
with section \ref{continua}.
Continuing
into the $c>0$ region, we have found that the solution gradually
changes its shape, with the hump in the  
imaginary part splitting into two (see Fig.\ref{conti}(a)).
At $c=0.7$ the branch turns back (Fig.\ref{biffy}).  
Note that the turning point occurs for $c$ much smaller
than $3{\tilde c}_2=1.04$, the saddle-node bifurcation point predicted
by the adiabatic analysis.
What is more important, the $\psi_+$ solution
turns not into the $\psi_-$ (as the adiabatic
approach predicted) but into some other
solution which has
two well-separated humps in the imaginary 
part.
 The  discrepancy is not surprising 
as the adiabatic approximation is valid only for small $c$,
where the shape of the solution is still well
reproduced by the single-hump constant-phase
 trial function (\ref{eq:ada-soln}).
 In the entire stretch between $c=0$
and the turning point, the solution remains stable.

Continuing this branch further, additional humps are added
as the solution passes a sequence of turning points.
Each pass of a
turning point results in the creation of a new hump in the
middle of the solitary wave. 
 As
we move along the branch, the distance between
successive turning points (the difference between
the corresponding values of $c$) becomes smaller and the 
new humps come with smaller amplitudes. As a result, 
a long plateau is formed which keeps on expanding as we continue
the branch (Fig.\ref{conti}(c)). The  broadening plateau
accounts for the vertical segment of the curve in Fig.\ref{biffy}.
If the bifurcation parameter $c$ is seen as a function
of the Sobolev norm $||\psi||$ (that is, if we turn Fig.\ref{biffy}
by $90^\circ$), the curve $c(||\psi||)$ has the form of a 
decaying oscillation.

The stability of the solution
alternates
 at each successive turning point. These changes 
  are due to a single real eigenvalue which
moves back and forth through the origin on the real line. 
(At the turning points the eigenvalue is right at the origin, of
course.)
The  lengths of the incursions this  eigenvalue 
makes into the positive and negative real lines decrease 
with each new turning point until the eigenvalue becomes indistinguishable
from zero.
Therefore the branch  becomes 
(neutrally) stable sufficiently `high up' in
$||\psi||$
in Fig.\ref{biffy} (i.e. for sufficiently long plateaus.)

Despite the fact that negative values of the
diffusivity $c$  are not
physically meaningful, 
 we did continue to  $c<0$ --- in the hope that the resulting 
branch would reach a turning point and then return to the positive
$c$-semiaxis (which it did indeed do).
The resulting branch is also shown in Fig.\ref{biffy}.
As we move into the region $c<0$, the $\psi_+$ 
solution gradually develops into
a three-humped state and when the curve returns  to $c=0$, we have
a complex of the
$\psi _{(-+-)}$ type, with a large 
separation between the individual solitons. (This complex of three 
solitons of
the parametrically driven nonlinear Schr\"{o}dinger equation was previously
found in Ref.\cite{BZ}.) Crossing into the $c>0$ region, 
the central $\psi_+$ soliton in the complex transforms as if it did 
not have the $\psi_-$ solitons attached to its flanks. As a result,
the $c>0$-portion of the corresponding $||\psi(c)||$ curve has virtually the 
same shape as the curve resulting from the 
continuation of $\psi_+$ to the region $c>0$; the only 
difference is that the curve emanating out of  $\psi _{(-+-)}$
is shifted upwards relative to the curve emanating out of $\psi_+$.
Similarly to the continuation of $\psi_+$, the 
continuation of $\psi _{(-+-)}$ goes via a
series of turning points, with each pass of the
turning point resulting in the creation of another hump in the
middle of the central region which 
becomes a long plateau. The lateral $\psi_-$
solitons are not affected by this process.
The linearised spectrum  is the union
of the spectrum of the long pulse described in 
the previous paragraph and spectra of two $\psi_-$ solitons.
In particular, it includes  two  positive real eigenvalues
contributed by the $\psi_-$'s and so the entire branch 
resulting from the continuation of $\psi _{(-+-)}$ is unstable.
We disregard it in what follows.

The plateau arising in the final stage of continuation of the 
two branches shown in Fig.\ref{biffy}, is nothing but an
interval on the $x$-axis where $\psi$ equals $\Psi_+^{(0)}$, 
the flat nonzero
solution given by Eq.(\ref{flat_back}). The 
corresponding value of $c$, $c_{\rm lim} =0.54$, 
falls within the  region $c \ge c_-$
where the background $\Psi_+^{(0)}$ is stable.
Here $c_-(\gamma, h)$ is given by Eq.(\ref{eq:cpmroots});
in particular,  $c_-(0.5,0.8)= 0.224$. 
In the language
of phase transitions, the long pulse shown in Fig.\ref{conti}(c)
can be seen as a ``bubble" of one stable phase
in another one.

\subsection{The bound state interpretation}
The long pulse can also be interpreted  as a bound state of two 
fronts interpolating between different stable backgrounds,
$\psi=0$ and $\psi=\Psi_+^{(0)}$.
This intuitively-appealing idea can be put on the quantitative 
footing by considering the 
spatial decay of perturbations to the flat background $\Psi_+^{(0)}$.

Indeed, consider the turning point separating the second stable branch
from the first unstable branch in Fig.\ref{biffy}. (This is 
a point with coordinates
$c=0.37$ and $||\psi||=2.48$. Note that 
we are only considering branches obtained
by the continuation of the soliton $\psi_+$ to
positive $c$. Branches obtained by the 
continuation of $\psi_-$ to negative $c$ first, are disregarded
here.) It is at this point that the 
modulus squared of $\psi(x)$ becomes double-humped; before that,
that is on the  branch that leads to this point (the first
unstable branch), the function $|\psi(x)|^2$ remained single-humped
despite the double-humped imaginary part.
Letting $h=0.8$, $\gamma=0.5$ and $c=0.37$
we check that $4A_+^2(A_+^2-1)> \gamma^2$ and hence, according to 
section \ref{spatially_uniform}, 
the quadratic (\ref{inequalityfork}), (\ref{eq:squad}) 
has two complex roots, $k^2=s_r + i s_i$ and $(k^2)^*$.
Therefore the wavenumber $k$ is complex as well:
  $k=k_r+ik_i$. Solving equation (\ref{eq:squad}) for $s$ we obtain,
  subsequently,
\begin{equation}
k_r=\frac{1}{\sqrt2} \left(s_r+\sqrt{s_r^2+s_i^2} \right)^{1/2}=
1.36, 
\quad 
k_i=\frac{s_i}{2k_r}=0.19.
\label{kk}
\end{equation}
(We have chosen positive values for 
$k_r$ and $k_i$ here.)

For $c$ near the turning point in question, the plateau has not yet 
formed between the two humps and so they can be crudely 
thought of as two overlapping NLS-like solitons with oscillations
on their adjacent (i.e. partner-facing) tails.
The bound state arises when one solitary
wave  is trapped 
in the potential well formed by the oscillatory
 tail of its partner. Making 
use of the potential of interaction of two attractive NLS solitons
\cite{Cai}, $U_{\rm eff} \propto \exp(-k_i z) \cos(k_r z)$
(where $z$ stands for the distance between the two humps), and taking
into account that $k_i/k_r \ll 1$, we obtain a rough estimate
 for the separation: $z=\pi/k_r$. This formula 
 could also be derived from purely qualitative considerations; 
 all one needs to notice is that the absolute value
 of the field $\psi$ obtained from the linear superposition 
 of two in-phase solitons is maximised (and so the energy is
 minimised) by placing one
 soliton
 at the first maximum of $|\psi|$ on the tail of its partner.  
 Substituting  $k_r$ from (\ref{kk}),
 the approximate formula gives $z=2.31$. This is in
qualitative agreement with the numerically found 
value $z=2.77$.

Moving further along the (second stable) branch, the plateau
appears and the two humps can no longer be approximated by
the NLS-like solitons. This makes
 the above estimate  invalid. 
 The solution can still be regarded as a bound state
 of two fronts but this time, in order to calculate 
 the characteristic width of the pulse one would need to know
 the full profile of the front.
 
We conjecture that 
for a given $h$ and $\gamma$, a free-standing 
stationary front exists just 
for a single value of $c$, namely $c=c_{\rm lim}$. (We are 
planning to verify this conjecture in future publications.)
On the other hand, stable and 
unstable bound states of fronts exist in  a finite interval of
$c$ values containing $c=c_{\rm lim}$ as an internal point.
It is fitting to note here that similar pulse-to-front 
transformations occur also in the other system
featuring the subcritical bifurcation, viz. the cubic-quintic Ginzburg-Landau
equation with internal gain \cite{Pulse_to_Front}.

\section{The $h$ vs $c$ diagram}
\label{h_vs_c}

As we mentioned in the previous section, there is a certain discrepancy
between the adiabatic analysis and numerical continuation. Numerically,
the
$\psi_{-}$ soliton was found to be continuable  all
the way to $c = + \infty$ whereas the adiabatic approach predicted the
existence of a turning point at $c = 3\tilde{c}_2$, where the
$\psi_{-}$ should have merged with the $\psi_{+}$ branch. 
As for the $\psi_+$
solution, we found that it turns into a 
pulse with the double-humped imaginary part 
(and not into the $\psi_{-}$ branch as suggested by the
adiabatic approximation.) In order to shed some light on the
possible source of the discrepancy
we performed the numerical
continuation of the pulse $\psi_{-}$ in $h$, for several fixed
values of $c$. Here, by the $\psi_{-}$ we mean the pulse solution which
results from the continuation of the Schr\"odinger $\psi_{-}$
soliton to positive $c$, for some fixed large value of $h$ (in our
case for $h = 0.8$.) Having obtained this starting-point solution for
several values of $c$, we then continued it in $h$, from $h = 0.8$
to smaller  $h$. The stability of the arising solutions was examined
by computing eigenvalues of the operator (\ref{EV_problem})
at sample values of $h$.

\begin{figure}
\includegraphics[ height = 2in, width =0.5\linewidth]{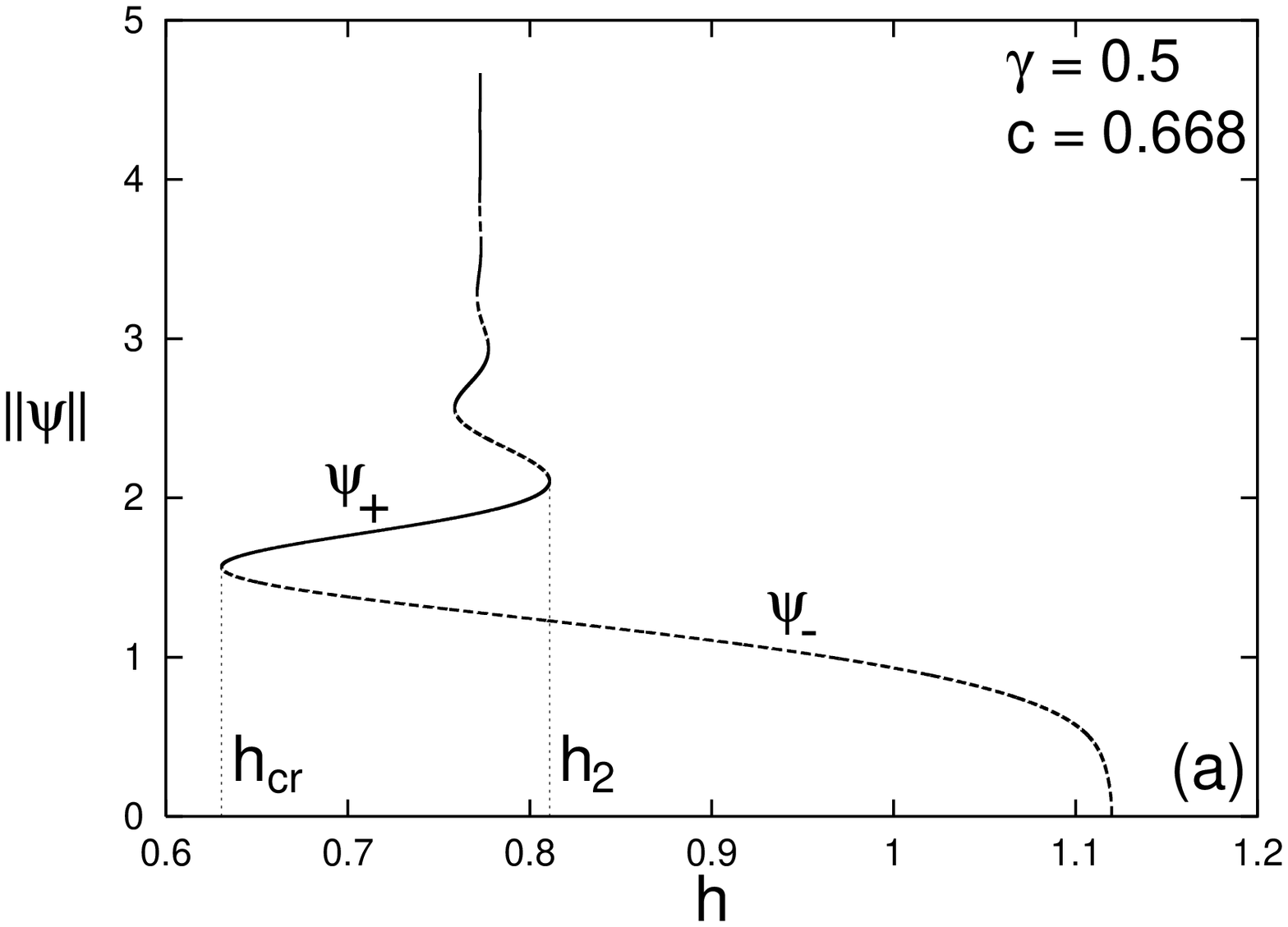}
\includegraphics[ height = 2in, width = 0.5\linewidth]{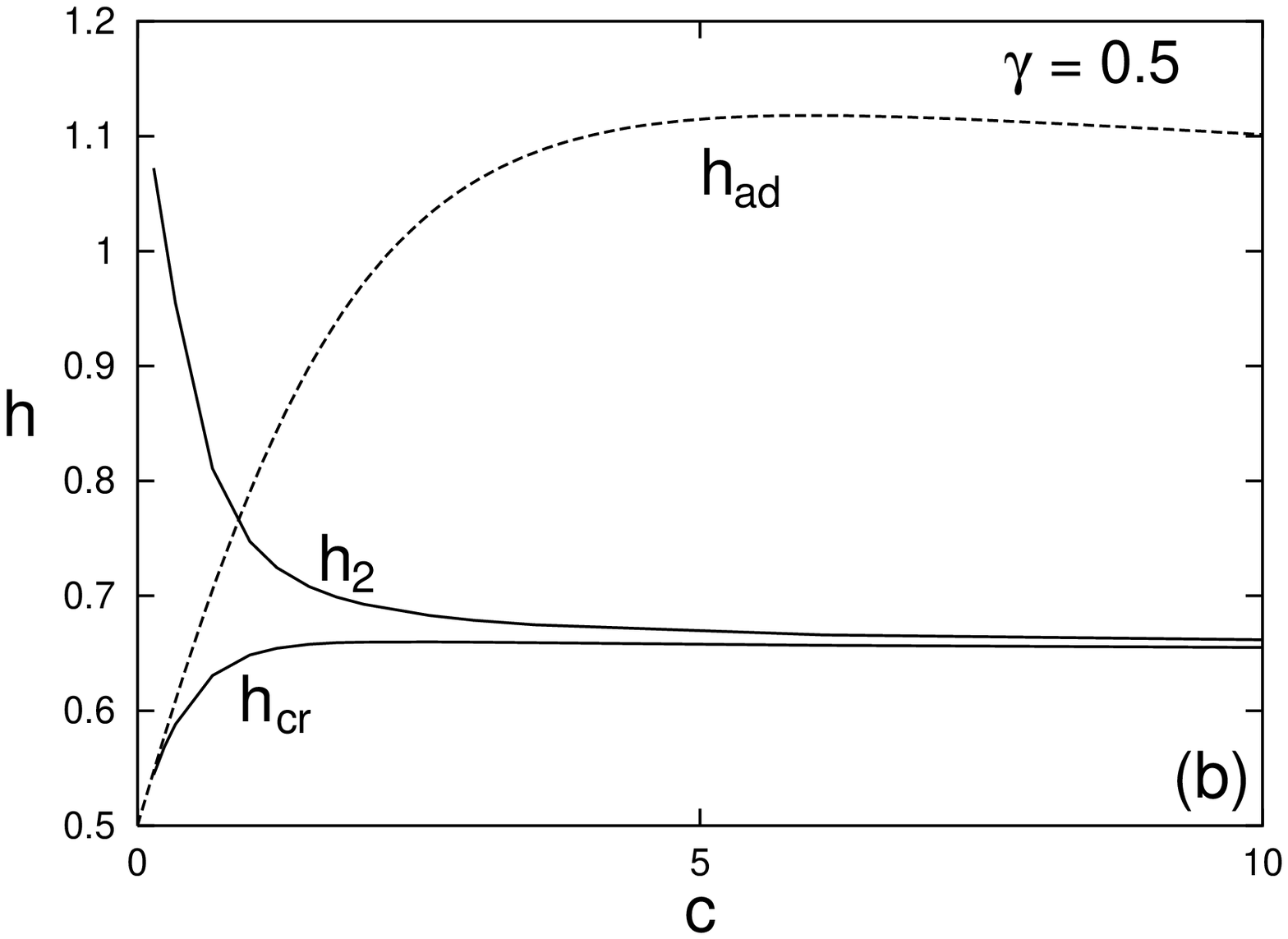}
\caption{\sf (a) The Sobolev norm of the 
solitary wave solution as a function of the driving
strength $h$. The solid and dashed lines indicate
stable and unstable branches, respectively.
 (b)
The  pulse existence region on the $(c, h)$-plane. 
Two pulse solutions, $\psi_{+}$ and $\psi_{-}$,
are born as $h$ exceeds the value $h_{\rm cr}(c)$ depicted by
 the lower solid line. The upper solid
curve gives the upper boundary of the $\psi_{+}$-pulse's
existence domain, $h_2(c)$. Also shown is the adiabatic approximation to the
saddle-node bifurcation curve, Eq.(\ref{threshold}) (dashed line).
}
\label{conti_in_h}
\end{figure}

In each case considered, the $\psi_{-}$ branch was found
to turn into the $\psi_{+}$ solution as $h$ reached the threshold
value $h_{\rm cr}=h_{\rm cr}(\gamma, c)$. (That is, for $h >
h_{\rm cr}$ there are two branches of solutions whereas for $h <
h_{\rm cr}$, there is none; see Fig.\ref{conti_in_h}(a)).
The entire $\psi_-$ branch is unstable; the single positive 
real eigenvalue moves to the negative semiaxis as the branch is 
continued past the turning point.
 Continuing the arising $\psi_{+}$ branch
to larger $h$, we reach another turning point at $h = h_2$, where
the $\psi_{+}$ solution transforms into a pulse with the double-humped
imaginary part.
The values $h_{\rm cr}$ and $h_2$ are shown in Fig.\ref{conti_in_h}(b), as
functions of $c$ (for the fixed $\gamma = 0.5$).
As $c \to \infty$, the difference
 $h_2-h_{\rm cr}$ decreases but remains nonzero. We verified this
 by computing $h_{\rm cr}$ and $h_2$ for equation
 (\ref{c_infty}) which pertains to $c=\infty$. In the same plot
 we display the function (\ref{threshold})  which gives the
adiabatic approximation to the curve 
 $h_{\rm cr} (c)$. (Note that for 
 small $c$, there is a good agreement between 
 numerical and approximate values  but as $c$
 grows, the two curves diverge.) 

Continuing the $\psi_{+}$ branch past the second turning point,
the solution adds another hump
in the middle of the pulse, turns back again, adds 
another one,
and so on. (See Fig.\ref{conti_in_h}(a)).
 A long plateau developes in the middle
of the pulse, just like when it was continued in $c$.
Similarly to the $c$-continuation, the turning points 
on the $h$-axis separate
regions of stability from regions of instability, with 
the instability being caused by a single positive real eigenvalue
(which moves through $\lambda=0$ at the turning points.)

From figure \ref{conti_in_h}(b) it is clear why the
saddle-node bifurcation point where the $\psi_{+}$ and $\psi_{-}$
solutions would merge, did not
appear in 
 Fig.\ref{biffy}. The reason  is that
  Fig.\ref{biffy} was plotted
for a relatively large value of $h$ $(h = 0.8)$ whereas
 according to Fig.\ref{conti_in_h}(b), a horisontal line $h = \mbox{const}$ 
 with $h>0.660$ 
 can have no intersections
with the saddle-node curve $h_{\rm cr}(c)$. (Here all the numbers are
for
$\gamma = 0.5$.)
The same  Fig.\ref{conti_in_h}(b) explains
what semed to be  a discrepancy between the adiabatic analysis
and the numerical continuation of the soliton
$\psi_-$ in $c$.  The numerical result that seemed
to contradict the adiabatics was that 
for $h=0.8$,  $\psi_-$  could be continued without bounds.
It is now obvious  from
 Fig.\ref{conti_in_h}(b)
that the unbounded continuation
is only possible  for $h$ greater than $0.660$.
  Continuing the $\psi_{-}$ soliton to positive $c$ for $h$ {\it
  smaller\/} than
  $0.660$, 
  the branch turns back (already as the $\psi_{+}$ pulse) after
hitting the lower solid curve in Fig.\ref{conti_in_h}(b).
Therefore, the  pattern arising for $h$ close 
to $\gamma$ actually {\it is\/} in qualitative agreement 
with the adiabatic
analysis, which was expected to be valid precisely for 
small $c$ or, equivalently, for small $h-\gamma$ differences.

%%%%%%%%%%%%%%%%%%%%%%%%%%%%%%%%%%%%%%%%%%%%%%%%%%%%%%%%%%%%%%%%%%%%%%%%%

\section{Concluding remarks}
\label{sec:conclusion}

In this paper we studied a cubic complex
Ginzburg-Landau equation in which linear losses and diffusion are
compensated by the linear parametric drive. The nonlinear term in
 the equation was taken to be purely conservative.

There are three stationary homogeneous solutions to equation
(\ref{eq:ddcgle}), and we have shown that the $\psi=0$ solution
is stable if $h \le \sqrt{1+ \gamma^2}$,
as long as $c \ge 0$.
This stability condition coincides with the corresponding condition
for the nonlinear Schr\"odinger case (i.e. for $c=0$).
On the other hand, the stability 
properties of the {\it nonzero\/} homogeneous solutions
are not the same as in the $c=0$ case.
 Indeed, unlike for $c=0$, there is 
 a stable   flat {\it nonzero\/} solution
$\psi=\Psi_+^{(0)}$  for sufficiently large
diffusion coefficients, $c \ge c_-(h, \gamma)$.

%Having established the persistence of the nonlinear
%Schr\"odinger solitons $\psi_+$ and $\psi_-$ for small nonzero $c$, we
%constructed the corresponding solutions in adiabatic approximation.
%The main result here is an analytical
%formula for the threshold driving strength, Eq.(\ref{threshold}).

Having established the persistence of the NLS 
solitons $\psi_+$ and $\psi_-$ for small nonzero $c$,
we continued them in $c$ numerically.
  The continuation of the soliton
$\psi_+$    yields a sequence of pulse-like solutions, separated
by turning points, with increasing number of humps. 
The stability of these solutions changes at each turning point, 
so that  stable multihump solutions coexist with unstable ones.
It is fitting to note here that the multistability of multipulse
solutions is {\it not\/} observed in the Schr\"odinger limit where
only the two-soliton complex was found to be stable \cite{BZ}.
As $c \to c_{\rm lim}$,  where $c_{\rm lim}=c_{\rm lim}(h, \gamma)$, 
the solution
takes the form of a long plateau (an interval of  the stable 
background  $\Psi_+^{(0)}$) sandwiched between two fronts.

We also performed the continuation in $h$, for the fixed $c$. For
each $c>0$, two localised solutions are born in a saddle-node 
bifurcation as $h$ exceeds a threshold value. (We obtained an analytic formula 
for the threshold in the adiabatic approximation;
it is in agreement with the numerical results for small $c$.)
 The subsequent continuation gives rise to a
sequence of coexisting stable multihump solutions culminating in
a bound state of two widely separated fronts.

Solitary pulses in the form of long shelves (plateaus) can 
allow easy experimental observation in physical systems described by 
our model. In particular,  they may be  employed as 
a natural basis for the
non-return-to-zero (NRZ) format of the data transmission in optical
telecommunications.  In the NRZ format, the 1
and 0 bits are coded, respectively, by sending or withholding 
 the signal
within a standard time slot. 

A string of several 1's
looks as a long uniform pulse of an essentially  arbitrary length. The
stability of such pulses is crucial to maintain the
fixed shape of the long array of 1's, and to prevent the 
{\it inter-symbol interference\/}, i.e., the blurring of
empty intervals between such strings, which represent (strings of) 0's
(see, e.g. Ref. \cite{ECI} and references therein.) 
 In the case
of lasers, which can also be described by the present model 
\cite{Longhi_GL}, the possibility of the generation of stable long pulses 
of arbitrary duration,
i.e., an effective \emph{tunability} of the output, is an essential
advantage too.

\acknowledgments

We thank Nora Alexeeva for her advice on numerics.
The first author's (I.B.'s) work was supported
 by the NRF
 of South Africa under grant No.2053723,
 by the Johnson Bequest Fund and the URC of the University of Cape Town.
The second author (S.C.) was  supported by the NRF 
and the Harry Crossley Foundation.
B.A.M. appreciates hospitality of the Department of
Physics at the Universit\"{a}t Erlangen-N\"{u}rnberg, and a
partial support from the European Science Foundation, through the 
``$\pi$-shift'' programme.

\end{document}